*Darwin et la socialité : entre anthropomorphisme et gradualisme.*

*Between anthropomorphism and gradualism: Darwin and sociality.*

Gérald FOURNIER[1]

**Abstract**

We propose, in this article, an analysis of the Darwin's approach to sociality. Sociality is perfectly integrated into the selective model, and is caused by the same process as struggle for existence. Thus, the selective process does not prohibit demonstrating cooperation, but on the contrary, explains it. We will see that the unification of sociality, in a more adequate representation to mammals, is not an expression of a naive anthropomorphism, but of a methodological anthropomorphism; normal consequence of his gradualist approach of phylogeny.

**Résumé**

On proposera dans cet article une analyse de l'approche de Darwin quant à la socialité. Cette dernière se trouve parfaitement intégrée dans le modèle sélectif, causée par le même processus de *lutte pour l'existence*. Le processus sélectif n'interdit donc en rien un résultat témoignant de la coopération, mais au contraire, l'explique. On verra que cette unification de la socialité, sous une représentation plus appropriée aux mammifères, n'est pas l'expression d'un anthropomorphisme naïf, mais d'un anthropomorphisme méthodologique ; conséquence normale de son approche gradualiste de la phylogenèse.

**Mots clés :** Instincts sociaux – sélection naturelle – socialité – gradualisme – anthropomorphisme – sympathie.

---

[1] Laboratoire d'Études du Phénomène Scientifique (LEPS), E.A. 4148, *Université de Lyon, Université Lyon 1*, France, Villeurbanne.

Le terme de socialité peut être utilisé pour embrasser le vaste ensemble hétérogène des modes de vies sociaux, du grégarisme à l'eusocialité évoluée, sans exclure la socialité humaine. C'est d'ailleurs par une perspective semblable que Darwin envisage le phénomène général du "vivre ensemble". Cette unification s'opère d'une part, en subsumant la socialité sous « l'instinct social » et, d'autre part, par une sympathie traversant les pôles de la taxonomie ; comme si Darwin défendait un certain monophylétisme de la socialité.

Dès *l'origine des espèces* (Chapitre VII) Darwin proposait une sélection d'instinct afin de rendre compte, par exemple, de comportements constructeurs à vocation collective. Dès cet ouvrage émergeait une entorse à "l'individualisme sélectif" (Gayon, 1992) sous la forme d'un principe *d'utilité "collective"* pour rendre compte de l'existence, malgré sa nocivité pour son porteur, de l'aiguillon barbelé de l'abeille. Mais c'est seulement avec *The Descent of Man*, qu'apparaissent les instincts sociaux.

Instincts sociaux, affections parentales et sociales, sympathie, lutte et principe d'utilité de groupe : tels sont les mots clés de la socialité chez Darwin. On traitera successivement les rapports entre l'instinct et l'intelligence, la nature des instincts sociaux, pour finir par la question des causes de l'anthropomorphisme darwinien de la socialité.

# 1. Instinct, sociabilité et intelligence

Dès *l'Origine des espèces*, Darwin souligne le rôle des instincts et les considère comme des caractères à part entière, transmis par l'hérédité, sujets à des variations et, dès lors, objets de sélection[2].

## 1.1) Gradation de l'instinct à l'intelligence

Si Darwin considère l'instinct d'une manière plutôt intuitive et floue, il range néanmoins celui-ci dans les « actes intellectuels distincts ». Par conséquent, il faut voir l'instinctif comme étant associé au système nerveux. Reste que Darwin ne veut pas pousser la définition, il en fait appel au sens commun. L'instinct se manifeste surtout dit-il lorsque l'animal est « jeune et sans expérience » ou, lorsque l'on observe « un acte accompli par beaucoup d'individus, de la même manière, sans qu'ils sachent en prévoir le but, alors que nous ne pourrions accomplir ce même acte qu'à l'aide de la réflexion et de la pratique »[3]. Ici, l'instinct *semble* s'opposer à la réflexion et à l'intelligence. Dans *The Descent of Man*, Darwin hésitera entre l'argument de la

---

[2] « Si l'on peut démontrer que les instincts varient si peu que ce soit, il n'y a aucune difficulté à admettre que la sélection naturelle puisse conserver et accumuler constamment les variations de l'instinct, aussi longtemps qu'elles sont profitables aux individus. » Darwin, *l'Origine des espèces*, GF, Paris, 1992, p. 263.

[3] Darwin, *l'Origine des espèces*, GF, Paris, 1992, p. 262.

« raison inverse » entre instinct et intelligence, qu'il prend de Cuvier, et une cohabitation entre instincts complexes et intelligence s'appuyant sur Pouchet.

L'hypothèse qu'apporte Darwin pour solutionner un gradualisme des facultés que tout semble opposer est la suivante :

> « Nous savons bien peu de choses sur les fonctions du cerveau, mais nous pouvons concevoir que, à mesure que les facultés intellectuelles se développent davantage, les diverses parties du cerveau doivent être en rapports de communications plus complexes, et que, comme conséquence, chaque portion distincte doit tendre à devenir moins apte à répondre d'une manière définie et héréditaire, c'est-à-dire instinctive, à des sensations particulières. »[4].

C'est pourquoi un même trait, – disons, comme la sympathie –, peut, en étant accompagné d'un développement cognitif suffisant, être modifié et socialisé, comme dans le cas de l'homme. Ainsi, à partir d'une base instinctive, comme l'instinct de sympathie, et avec le développement graduel des facultés intellectuelles, la sympathie peut se trouver transformée selon l'environnement social et sa perception de l'utilité à un instant *t*. La maturation est progressive et c'est l'approche gradualiste qui explique comment Darwin peut parler tantôt de la sympathie comme instinct altruiste, tantôt comme un socle configuré par la société, donc ouvert à l'apprentissage. C'est là un début de réponse au philosophe qui, par ses habitudes théoriques, sait bien que le propre d'un instinct est justement de ne pas fonctionner selon la plasticité ; qu'un instinct ne saurait devenir une faculté – ici de contagion des normes, par intériorisation et ré-émission des éloges et des blâmes d'autrui.

Or, Darwin ne suivrait probablement pas ce raisonnement pourtant si juste en apparence. Il nous inviterait plutôt à sortir de notre "caverne" biologique, conceptuelle, philosophique et religieuse ; celle qui nous pousse à toujours raisonner dans cette temporalité existentielle étroite, toujours à partir d'ici et maintenant, amplifiant la différence anthropologique et la condition – figée, bien entendu – des vivants non-humains. Comme si le monde demeurait toujours le même. Comme si le cognitif, l'apprentissage et la technique étaient nés avec *l'homo sapiens* ; pire, une invention et un monopole de l'humain. C'est cette temporalité là qui nourrit la perception de notre "étrangeté" toute gnostique dans le monde, avec tout ce que cela implique quant à notre façon de considérer la nature et le vivant. Réintégrer la temporalité évolutive dans les choses les anoblit en nous montrant que chaque forme de vie est « résolution de problèmes », comme le dirait K. Popper.

Darwin ne peut opposer radicalement instincts et intelligence car ils ont tout deux, nécessairement, une origine commune par et dans le système nerveux. On le voit, la logique évolutive vient déstructurer les oppositions métaphysiques et toutes langagières qui continuent à maintenir en nous leurs suprématies sur la réalité, comme si de rien n'était. Or, n'est-il pas évident qu'une psychologie de la connaissance se doit de prendre en compte cette temporalité étendue dans laquelle s'est construit notre système bio-cognitif ? Darwin n'était-il pas conscient de ce rôle de la théorie évolutive lorsqu'il écrivit dans ses *Carnets* : « Celui qui comprend le Babouin en fera plus pour la métaphysique que Locke »[5] ?

---

[4] Darwin, *La Descendance de l'Homme et la Sélection Sexuelle*, (1871) Schleicher Frères, 2ᵈᵉ éd. (1874), trad. fr. par Edmond Barbier, ouvrage publié sans date (660 pages + 38 planches, 1906 ?), pp. 69-70.

Il apparaît ainsi une ambivalence délicate dans l'analyse des rapports entre instinct et intelligence, entre raison inverse, gradation et cohabitation. Ce qui est développé chez Darwin n'est donc pas seulement une sociobiologie, comme l'insufflera Wilson, mais aussi une psychologie évolutionniste[6]. Si Darwin reste prudent quant à la « raison inverse » entre instinct et intelligence, celle entre la force physique et la socialité est en revanche plus tranchée.

**1.2) La force est en raison inverse avec la sociabilité**

Il est intéressant de noter qu'avant de s'engager sur la sociogenèse, Darwin indique une opposition entre la force physique d'un animal et le mode de vie social. Il estime ainsi qu'il est « peu probable qu'un animal de grande taille, fort et féroce, et pouvant, comme le gorille, se défendre contre tous ses ennemis, puisse devenir un animal sociable ; or ce défaut de sociabilité aurait certainement entravé chez l'homme le développement de ses qualités mentales d'ordre élevé, telle que la sympathie et l'affection pour ses semblables. Il y aurait donc eu, sous ce rapport, un immense avantage pour l'homme à devoir son origine à un être comparativement plus faible. »[7]. Rajoutant un peu plus loin que : « les premiers ancêtres de l'homme étaient sans doute inférieurs, sous le rapport de l'intelligence et probablement des dispositions sociales, (...) mais on comprend parfaitement qu'ils puissent avoir existé et même prospéré, si tandis qu'ils perdaient peu à peu leur force brutale et leurs aptitudes animales, telles que celle de grimper sur les arbres, etc., ils avançaient en même temps en intelligence. »[8].

À la possibilité de la raison inverse entre instinct et intelligence, que Darwin reprend de Cuvier, notre savant conçoit un rapport analogue entre la force physique et la sociabilité. Il semble que Darwin raisonne ici en termes d'usage, comme c'est le cas touchant les rapports entre la libération de la main et la bipédie[9], dans une perspective de *"new deal"* anatomique que l'effet de l'usage et du non-usage facilite et implique. La faiblesse physique est à la fois condition et résultat de la socialité. L'idée d'une redistribution du "budget" anatomique propre à la composante sociale du biotope ou *"sociotope"* semble être ici considérée ; idée confirmée ultérieurement par l'état de civilisation où priment alors les facultés mentales et morales. Mais tout cela concerne *l'état* social et ne répond pas à la question du passage du mode de vie solitaire au mode de vie social (sociogenèse). Comment et sur quoi Darwin articule ce passage, c'est ce que l'on se propose d'examiner maintenant.

---

[5] "He who understands baboon would do more toward metaphysics than Locke". Darwin, *Notebook M*, 1838.

[6] « J'entrevois dans un avenir éloigné des routes ouvertes à des recherches encore bien plus importantes. La psychologie sera solidement établie sur une nouvelle base, c'est-à-dire sur l'acquisition nécessairement graduelle de toutes les facultés et de toutes les aptitudes mentales. » Darwin, *l'Origine des espèces*, GF, Paris, 1992, p. 547.

[7] Darwin, *La Descendance de l'Homme*, pp. 64-65.

[8] *Ibid.*, p. 65.

[9] *Ibid.*, pp. 51-53.

## 2. Fonctionnement, origine et spécificité des instincts sociaux

### 2.1) L'instinct social fonctionne, comme les autres instincts, par le plaisir et la peine

Darwin reste, quant à la sociogenèse, dans une perspective relative à l'intérêt individuel. Cet intérêt se manifeste par des sentiments. On trouve dans le texte un renversement de l'optique théorique entre affects et socialité, Darwin soutenant que si l'on « a souvent affirmé que les animaux sont d'abord devenus sociables, et que, en conséquence, ils éprouvent du chagrin lorsqu'ils sont séparés les uns des autres, et ressentent de la joie lorsqu'ils sont réunis ; il est bien plus probable que ces sensations se sont développées les premières, pour déterminer les animaux qui pouvaient tirer un parti avantageux de la vie en société à s'associer les uns aux autres ; de même que le sentiment de la faim et le plaisir de manger ont été acquis d'abord pour engager les animaux à se nourrir. »[10]. L'instinct que Darwin appelle « social » fonctionnerait ainsi comme les instincts vitaux, accompagnés de plaisirs et de peines qu'ils sont, comme en témoigne encore le passage suivant : « quant à l'impulsion, qui conduit certains animaux à s'associer et à s'entr'aider de diverses manières, nous pouvons conclure que, dans la plupart des cas, ils sont poussés par les mêmes sentiments de joie et de plaisir que leur procure la satisfaction d'autres actions instinctives, ou par le sentiment de regret que l'instinct non satisfait laisse toujours après lui. »[11]. Pas d'altruisme réciproque : ce sont les instincts sociaux, sélectionnés naturellement, qui poussent littéralement les différents individus à l'agrégation. L'instinct social fonctionne donc comme les autres instincts, selon le plaisir et la peine, bien qu'il ait comme spécificité d'être, comme on va le voir, *toujours présent et persistant*. On retient donc que la socialité est, chez Darwin, précédée d'instincts sociaux (objets de sélection) se manifestant sous la forme de plaisirs relatifs au "vivre ensemble" (effets psychologiques causant l'attraction sociale).

### 2.2) Les instincts sociaux sont d'origine filiale

Cependant, l'on ne saurait marquer le passage dans la voie sociale, par la simple expression « instinct social ». Il faut une attache plus ancienne d'où l'on puisse faire dériver ce nouveau type d'instinct. D'où dérive l'affection sociale ? Darwin répond, avec un sentiment d'évidence :

> « L'impression de plaisir que procure la société est probablement une extension des affections de parenté ou des affections filiales ; on peut attribuer cette extension principalement à la sélection naturelle, et peut-être aussi, en partie, à l'habitude. Car, chez les animaux pour lesquels la vie sociale est avantageuse, les individus qui trouvent le plus de plaisir à être réunis peuvent le mieux échapper à divers dangers (…). Il est inutile de spéculer sur l'origine de l'affection des parents pour leurs enfants et de ceux-ci pour leurs parents ; ces affections constituent évidemment la base des affections sociales »[12].

La gradation de l'affection parentale à l'affection sociale aurait ainsi pour origine une variation de l'instinct poussant divers individus à entrer plus fréquemment en interaction avec leurs congénères. Cette légère variation, en donnant un avantage dans la lutte pour la vie, aurait ainsi été sélectionnée et généralisée. On reconnaît ici le primat, l'orthodoxie explicative, de l'individualisme sélectif (Gayon, 1992). La socialité part de l'intérêt individuel. Elle est fondée instinctivement et se manifeste par un sentiment de plaisir dans la subjectivité de l'individu. Ce qui est d'autant plus intéressant, et qu'il faut exposer maintenant, c'est le rôle de ces instincts dans la possibilité morale.

---

[10] Darwin, *La Descendance de l'homme*, p. 112.
[11] *Ibid.*, p. 111.
[12] *Ibid.*, pp. 112-113.

### 2.3) Les instincts sociaux sont toujours présents et persistants

Le propre des instincts sociaux, qui explique pourquoi ils peuvent être en lutte avec les autres instincts est, si l'on veut, dans le langage psychologique, qu'ils ne se "déchargent" pas. C'est pourquoi ils peuvent et vont servir de référentiels, notamment concernant le sens moral humain[13]. Darwin illustre ce conflit d'instincts, chez l'homme, de la manière suivante : « lorsqu'un désir, lorsqu'une passion temporaire l'emporte sur ses instincts sociaux, il réfléchit, il compare les impressions maintenant affaiblies de ces impulsions passées avec l'instinct social toujours présent, et il éprouve alors ce sentiment de mécontentement que laissent après eux tous les instincts auxquels on n'a pas obéi. Il prend en conséquence la résolution d'agir différemment à l'avenir ; – c'est là ce qui constitue la conscience. »[14]. Il est aussi possible de faire référence à un autre passage qui fait penser à quelque chose de l'ordre de la différence kantienne[15] entre agir *par devoir* et agir *conformément au devoir*, même si, bien entendu, le référentiel est ici fondé naturellement et non rationnellement : « un désir ou un instinct peut pousser un homme à accomplir un acte contraire au bien d'autrui ; si ce désir lui paraît encore, lorsqu'il se le rappelle, aussi vif ou plus vif que son instinct social, il n'éprouve aucun regret d'y avoir cédé ; mais il a conscience que, si sa conduite était connue de ses semblables, elle serait désapprouvée par eux, et il est peu d'hommes qui soient assez dépourvus de sympathie pour n'être pas désagréablement affectés par cette idée. S'il n'éprouve pas de pareils sentiments de sympathie, si les désirs qui le poussent à de mauvaises actions sont très énergiques à certains moments, si, enfin, quand il les examine froidement, ses désirs ne sont pas maîtrisés par les instincts sociaux persistants, c'est alors un homme essentiellement méchant ; il n'est plus retenu que par la crainte du châtiment et la conviction qu'à la longue il vaut mieux, même dans son propre intérêt, respecter le bien des autres que consulter uniquement son égoïsme. »[16].

Les instincts sociaux jouent donc un rôle tout à fait central dans la *possibilité* morale. Il s'agit certes, d'une morale dérivée de l'utilité, mais cela n'empêche qu'elle n'en incarne pas moins le souci d'autrui et de la communauté. Par les instincts sociaux et leur sélection, l'homme possède une nature sociale et morale. Et c'est précisément cela qui autorise Darwin à conclure : « ainsi se trouve écarté le reproche de placer dans le vil principe de l'égoïsme les bases de ce que notre nature a de plus noble ; à moins, cependant, qu'on appelle égoïsme la satisfaction que tout animal éprouve lorsqu'il obéit à ses propres instincts, et le regret qu'il ressent lorsqu'il en est empêché. »[17]. Grace aux instincts sociaux, l'individu a en lui un mobile biologique (égoïste) d'action altruiste (pour la communauté).

---

[13] Sympathie, *instincts sociaux persistants* et mémoire sont, chez Darwin, les traits indispensables au sens moral ou conscience.
[14] Darwin, *La Descendance de l'homme,* p. 642.
[15] Darwin, parlant du devoir, fait d'ailleurs référence à Kant dès le début du chapitre IV et entend bien répondre à la question de l'origine de ce puissant principe.
[16] Darwin, *La Descendance de l'homme,* pp. 124-125.

Tous les instincts et tous les comportements sont donc soumis à variation et à sélection. Les instincts sociaux ne font pas exception. Avec ces derniers, le plaisir de l'individu et l'intérêt de la communauté sont convergents. Par contre, une chose saute aux yeux : le traitement unifié de la socialité sur un modèle que nous attribuerions aujourd'hui aux mammifères. C'est qu'il y a clairement un *lien* au sens de Lorenz, c'est-à-dire une reconnaissance individuelle et une affectivité dans l'approche darwinienne de la socialité. Or, cela est contraire à l'ordre de la phylogenèse que de mettre l'affectivité en premier, alors qu'elle exige un certain raffinement évolutif. On retrouve par exemple notre auteur soulignant que les fourmis « reconnaissent leurs camarades après plusieurs mois d'absence et éprouvent de la sympathie les unes pour les autres. »[18]. On le voit, les mécanismes amenant à la socialité sont censés être identiques. La composante affective y est inévitable ; c'est elle qui est l'objet et l'effet premier de sélection (sur l'individu). Le scientifique sait que ce type de socialité est en vérité bien difficilement comparable avec l'eusocialité évoluée, son polyéthisme de caste, son polymorphisme et son *intelligence collective simple*. Il s'agit alors de rendre raison de cet anthropomorphisme un peu inattendu.

## 3. Entre anthropomorphisme et gradualisme

L'historien des sciences peut être surpris d'un certain retour à l'anthropomorphisme, là où Buffon, en bon cartésien, ne fautait pas. Pour lui, ces sociétés d'insectes ne constituent qu'un « assemblage physique ordonné par la nature, et indépendant de toute vue, de toute connaissance et de tout raisonnement »[19]. Et cette intelligence n'est que collective, « qu'un résultat purement mécanique, une combinaison de mouvement proportionnelle au nombre. Leur intelligence apparente ne vient que de la multitude réunie »[20].

### 3.1) Anthropomorphisme et sélection naturelle

L'accusation d'anthropomorphisme au sein des travaux de Darwin touche généralement son hypothèse de la sélection naturelle et même, plus précisément, l'expression même de « sélection ». Darwin répondait à l'objection par un argument d'ordre heuristique[21]. Or, ce terme est précisément en contradiction

---

[17] *Ibid.*, p. 130.

[18] *Ibid.*, p. 160.

[19] Buffon, G.-L. Leclerc, Daubenton, L.-J.-M., 1753, *Histoire naturelle générale et particulière*, tome IV, Imprimerie royale, Paris, 541, p. 94.

[20] *Ibid.*, p. 93.

[21] « Dans le sens littéral du mot, il n'est pas douteux que la sélection naturelle est un terme erroné ; mais qui donc

directe avec la vision darwinienne du monde. Une vision, qui, justement, nous apprend l'existence d'une toute autre intelligence créatrice, bien loin de nos procédés de cognition (intention-action) – ou de ce que Dennett appelle la théorie de *l'esprit-premier*[22]. En comparaison, la sélection artificielle et la sélection sexuelle sont de bonnes expressions, car il y a alors une cognition agissante, une intentionnalité. Alors que le propre de la sélection naturelle est justement cette auto-organisation créatrice dépourvue de toute planification cognitive commensurable avec la cognition humaine. Le mécanisme est donc certes comparable, mais la cause motrice ou efficiente est bien différente, si du moins l'on s'accorde à penser que la négation argumentée de *l'esprit-premier* est un résultat philosophique important du travail de Darwin. Cette critique se retrouve dans le livre de Daniel C. Dennett, *Darwin's Dangerous Idea. Evolution and the Meanings of Life*[23]. L'anthropomorphisme de la socialité est quant à lui bien moins condamnable.

### 3.2) Un anthropomorphisme cognitif et social issu du gradualisme

Darwin, et ce point est significatif, s'en tient à cette considération d'une socialité de type mammifère en faisant référence à P. Huber[24], qui aurait « clairement démontré que les fourmis peuvent, après une séparation de quatre mois, reconnaître leurs camarades appartenant à la même communauté »[25].

Il est donc déjà possible de soutenir que la perception darwinienne de la socialité n'est pas une faiblesse anthropocentriste, mais s'appuie sur les travaux et les représentations scientifiques de l'époque. C'est ainsi que Darwin attribue aux animaux des affects, de la sympathie au jeu. En effet, dit-il, « tous les animaux éprouvent de l'étonnement, et beaucoup font preuve de curiosité »[26] et « les insectes eux-mêmes jouent les uns avec les autres »[27]. Parlant du processus général de la cognition, et en particulier de celui de l'association d'idées, Darwin conclut que « dans tous les cas, c'est une pure supposition que d'affirmer que

---

a jamais reproché aux chimistes de parler des affinités électives des divers éléments ? - Or, on ne peut pas dire, à strictement parler, que l'acide choisisse la base avec laquelle il se combine de préférence. On a dit que je parle de la sélection naturelle comme d'un pouvoir actif ou comme une Divinité ; mais qui reproche à un auteur de parler de l'attraction ou de la gravitation comme gouvernant les mouvements des planètes ? Chacun sait ce que signifient et ce qu'impliquent ces expressions métaphoriques ; elles sont quasiment nécessaires à la concision de l'expression. » Darwin, *L'Origine des espèces,* troisième édition, (1861), 1959, p. 165.

[22] Plus qu'une théorie, c'est une illusion naturelle de notre espèce, qui comprend, comme on pourrait le dire aujourd'hui, en mobilisant sa *théorie de l'esprit* sur les choses. Or, *la théorie de l'esprit* est adaptée au seul déchiffrement des intentions d'autrui. Quant à la théorie de l'esprit-premier, c'est une théorie-illusion qui met l'esprit, soit le plus complexe et le plus tardif évolutivement, au début de l'univers. C'est bien mettre au début ce qu'il y a à la fin.

[23] « Ce fait fonde l'autre famille de métaphores qui a à la fois stimulé et entravé, éclairé et égaré les penseurs qui se sont trouvés confrontés à l'"étrange inversion de raisonnement" de Darwin : l'attribution apparente d'intelligence au processus même de sélection naturelle dont Darwin soulignait qu'il n'est pas intelligent. N'était-ce pas malheureux, en fait que Darwin ait choisi d'appeler son principe "*sélection* naturelle", avec tout ce que cela implique de connotations anthropomorphiques ? (…) Bien des gens se fourvoyèrent, et Darwin était enclin à s'en attribuer la responsabilité : "Je dois m'être mal expliqué", dit-il, concédant : "Je suppose que « sélection naturelle » n'était pas le terme approprié" (Desmond et Moore, 1991, p. 492). » Dennett, *Darwin est-il dangereux ?*, Odile Jacob, Paris, 2000, p. 83.

[24] P. Huber, *Les Mœurs des fourmis*, 1810, p. 150.

[25] Darwin, *La Descendance de l'homme*, p. 77.

[26] *Ibid.*, p. 73.

[27] *Ibid.*, p. 70.

l'acte mental n'a pas exactement la même nature chez l'animal et chez l'homme. Si l'un et l'autre rattachent ce qu'ils conçoivent, au moyen de leurs sens, à une conception mentale, tous deux agissent de la même manière. »**28**. C'est exactement dans la même perspective évolutive gradualiste que Darwin estimait que les mécanismes de la socialité devaient être sensiblement les mêmes. Et, face à l'objection d'anthropomorphisme, il est probable que notre savant aurait mobilisé la considération du cerveau des fourmis, « ce merveilleux atome de matière » :

> « Le cerveau a certainement augmenté en volume à mesure que les diverses facultés mentales se sont développées. Personne, je suppose, ne doute que, chez l'homme, le volume du cerveau, relativement à celui de son corps, si on compare ces proportions à celles qui existent chez le gorille ou chez l'orang, ne se rattache intimement à ses facultés mentales élevées. Nous observons des faits analogues chez des insectes : chez les fourmis, en effet, les ganglions cérébraux atteignent une dimension extraordinaire (...) le cerveau d'une fourmi est un des plus merveilleux atomes de matière qu'on puisse concevoir, peut-être même plus merveilleux encore que le cerveau de l'homme. »**29**.

Sa perspective d'unification du vivant sous le fonctionnement sélectif, son gradualisme enfin, lui interdisaient de trop segmenter les choses ou d'œuvrer pour ce qu'il, en connaissance de cause, devait combattre : une pensée de la rupture et de surélévation, voire d'isolement, anthropologique.

**Conclusion :**

Il ne faut pas confondre, dans l'œuvre de Darwin, ce que l'on pourrait appeler une *thèse générale* (celle de la « descendance avec modification » ou évolution) avec celle d'une *thèse spéciale* comme l'est le gradualisme. Ce n'est pas parce que le gradualisme semble invalidé que la thèse générale l'est, et ce, pour la seule raison que Darwin considérait cette thèse spéciale comme vitale à sa théorie. De même, plusieurs thèses concernant la socialité peuvent être distinguées :

- une *thèse fondamentale* concevant le comportement comme objet de sélection. C'est là, pour Darwin, une condition nécessaire à l'intégration de la socialité dans le modèle sélectif ;
- une *thèse générale* qui en découle : celle d'une *sélection d'instincts dits "sociaux"*, soutenant que la socialité est objet de sélection, tant dans sa naissance que dans son extension (par *lutte* de groupe) ;
- une *thèse spéciale* : celle d'une *socialité à base affective* qui accompagne celle d'une *socialité d'origine filiale*.

Si la thèse d'une socialité unifiée par l'hypothèse de son origine affective et familiale, accompagnée du prisme déformant d'une sympathie méta-spécifique, peut être perçue, rétrospectivement, comme la projection d'une socialité de type "mammifère" sur l'ensemble de la socialité, il nous a paru adéquat de défendre qu'il s'agissait là d'une conséquence normale de l'approche gradualiste de Darwin. Ce dernier avait des raisons d'émettre ces hypothèses et était partiellement dans le faux pour de très bonnes raisons. Il est, certes, aisé de voir que l'eusocialité des éthologues et entomologistes contemporains est par principe étrangère à Darwin**30**. Il y a ainsi, chez Darwin, l'exigence d'une socialité unifiée qui s'oppose au

---

**28** *Ibid.*, p. 87.
**29** *Ibid.*, pp. 53-54.
**30** Pour prendre un exemple témoignant de la façon dont on pense aujourd'hui la socialité évoluée des insectes, il polyphylétisme de la socialité, avec ses nombreuses trajectoires possibles pour atteindre un même état social (Jaisson, 1985). Le problème est, en effet, qu'il n'y a pas de "phylogenèse" de la socialité (monophylétisme) ou de la cognition comme c'est le cas pour les espèces. Il y a plusieurs "foyers" de socialité (polyphylétisme) et de cognition qui répondent à différents types d'habitats et de niches, à différentes trajectoires évolutives. C'est en tous cas ce que semblent suggérer les connaissances actuelles.

Le gradualisme de notre savant, sa volonté d'insister sur l'attache entre les espèces, tout cela déborda nécessairement sur la socialité où les différences de degrés sont venues effacer des différences pourtant bien réelles. Cela n'empêche que ses thèses relatives à la sélection du comportement, à ce qui, en nous, est dédié au social, (comme l'instinct/faculté de sympathie), connaissent une certaine postérité. Les découvertes des "neurones miroirs" (Gallese et Rizzolatti, 1996)[31], de l'imitation néonatale (Meltzoff, 1977), de la fonction empathique de la zone préfrontale ventro-médiane (Damasio, 1994), ainsi que la proposition d'une *théorie de l'esprit* (Premack, 1978), l'attestent. Il semble toutefois que ce soit *L'expression des émotions* et non *La descendance de l'homme* qui ait retenu l'attention sur ces points.

---

est possible de mobiliser le travail de Grassé à qui l'on doit le concept de stigmergie : la construction s'auto-organise et appelle les stades ultérieurs de construction, sans plan extérieur. De même c'est la densité (le gradient) de phéromone diffusé par la reine qui, chez les Termites, selon un certain seuil, détermine la dimension de la loge (Theraulaz et Bonabeau, 1997). Enfin, les phéromones d'agrégation et les phénomènes d'intelligence collective, viennent témoigner d'un fondement et d'un fonctionnement social bien différent de ce qu'imaginait Darwin quant à l'objet eusocial.

[31] Cf. Rizzolatti (G) et Sinigaglia (C), *Les Neurones Miroirs*, Odile Jacob, Mayenne, 2008.